\documentclass[aps,twocolumn,amssymb,amsfonts,amsmath,showpacs,final,a4paper,superscriptaddress]{revtex4-2}

\usepackage{float}
\usepackage[dvips,final]{graphicx}
\usepackage{dcolumn}
\usepackage{bm}
\usepackage[latin1]{inputenc}
\usepackage{textcomp}
\usepackage{amsmath}
\usepackage{color}
\usepackage[normalem]{ulem}
\usepackage{hyperref} 
\hypersetup{colorlinks,linkcolor={blue},citecolor={blue},urlcolor={blue}}

\begin{document}

\title{Photo-induced charge-transfer renormalization in NiO}

\author{Tobias Lojewski}
 \affiliation{Faculty of Physics and Center for Nanointegration Duisburg-Essen (CENIDE), University of Duisburg-Essen, Lotharstr.~1, 47057 Duisburg, Germany}

\author{Denis Gole\v z}
 \affiliation{Jozef Stefan Institute, Jamova 39, 1000 Ljubljana, Slovenia}
 \affiliation{Faculty of Mathematics and Physics, University of Ljubljana, Jadranska 19, 1000 Ljubljana, Slovenia}

\author{Katharina Ollefs}
 \affiliation{Faculty of Physics and Center for Nanointegration Duisburg-Essen (CENIDE), University of Duisburg-Essen, Lotharstr.~1, 47057 Duisburg, Germany}

\author{Lo\"ic Le Guyader}
 \affiliation{European XFEL, Holzkoppel 4, 22869 Schenefeld, Germany}

\author{Lea K\"ammerer}
 \affiliation{Faculty of Physics and Center for Nanointegration Duisburg-Essen (CENIDE), University of Duisburg-Essen, Lotharstr.~1, 47057 Duisburg, Germany}

\author{Nico Rothenbach}
 \affiliation{Faculty of Physics and Center for Nanointegration Duisburg-Essen (CENIDE), University of Duisburg-Essen, Lotharstr.~1, 47057 Duisburg, Germany}

\author{Robin Y. Engel}
 \affiliation{Deutsches Elektronen-Synchrotron DESY, Notkestr. 85, 22607 Hamburg, Germany}

\author{Piter S. Miedema}
 \affiliation{Deutsches Elektronen-Synchrotron DESY, Notkestr. 85, 22607 Hamburg, Germany}

\author{Martin Beye}
 \affiliation{Deutsches Elektronen-Synchrotron DESY, Notkestr. 85, 22607 Hamburg, Germany}
 \affiliation{Department of Physics, AlbaNova University Center, Stockholm University, SE-10691 Stockholm, Sweden}

\author{Gheorghe S. Chiuzb\u{a}ian}
 \affiliation{Sorbonne Universit\'{e}, CNRS, Laboratoire de Chimie Physique - Mati$\grave{e}$re et Rayonnement, 75005 Paris, France} 

\author{Robert Carley}
\affiliation{European XFEL, Holzkoppel 4, 22869 Schenefeld, Germany}

\author{Rafael Gort}
\affiliation{European XFEL, Holzkoppel 4, 22869 Schenefeld, Germany}

\author{Benjamin E. Van Kuiken}
\affiliation{European XFEL, Holzkoppel 4, 22869 Schenefeld, Germany}

\author{Giuseppe Mercurio}
\affiliation{European XFEL, Holzkoppel 4, 22869 Schenefeld, Germany}

\author{Justina Schlappa}
\affiliation{European XFEL, Holzkoppel 4, 22869 Schenefeld, Germany}

\author{Alexander Yaroslavtsev}
\affiliation{European XFEL, Holzkoppel 4, 22869 Schenefeld, Germany}
\affiliation{Department of Physics and Astronomy, Uppsala University, 75120 Uppsala, Sweden}

\author{Andreas Scherz}
\affiliation{European XFEL, Holzkoppel 4, 22869 Schenefeld, Germany}

\author{Florian D\"{o}ring}
\affiliation{Paul Scherrer Institut, Forschungsstr. 111, 5232 Villigen PSI, Switzerland}

\author{Christian David}
\affiliation{Paul Scherrer Institut, Forschungsstr. 111, 5232 Villigen PSI, Switzerland}

\author{Heiko Wende}
 \affiliation{Faculty of Physics and Center for Nanointegration Duisburg-Essen (CENIDE), University of Duisburg-Essen, Lotharstr.~1, 47057 Duisburg, Germany}

\author{Uwe Bovensiepen}
 \affiliation{Faculty of Physics and Center for Nanointegration Duisburg-Essen (CENIDE), University of Duisburg-Essen, Lotharstr.~1, 47057 Duisburg, Germany}
 \affiliation{Institute for Solid State Physics, The University of Tokyo, Kashiwa, Chiba 277-8581, Japan}

\author{Martin Eckstein}
 \affiliation{I. Institute of Theoretical Physics, University of Hamburg, 20355 Hamburg, Germany}
 
\author{Philipp Werner}
 \affiliation{Department of Physics, University of Fribourg, 1700 Fribourg, Switzerland}

\author{Andrea Eschenlohr}
 \email[]{andrea.eschenlohr@uni-due.de}
 \affiliation{Faculty of Physics and Center for Nanointegration Duisburg-Essen (CENIDE), University of Duisburg-Essen, Lotharstr.~1, 47057 Duisburg, Germany}

\date{\today}%

\begin{abstract}
Photo-doped states in strongly correlated charge transfer insulators are characterized by  $d$-$d$ and $d$-$p$ interactions and the resulting intertwined dynamics of charge excitations and local multiplets. Here we use femtosecond x-ray absorption spectroscopy in combination with dynamical mean-field theory to disentangle these contributions in NiO. Upon resonant optical excitation across the charge transfer gap, the Ni $L_3$ and O~$K$ absorption edges red-shift for $>10$~ps, associated with photo-induced changes in the screening environment. 
An additional signature below the Ni $L_3$ edge is identified for $<1$~ps, reflecting a transient nonthermal population of local many-body multiplets. We employ a nonthermal generalization of the multiplet ligand field theory to show that the feature originates from $d$-$d$ transitions. Overall, the photo-doped state differs significantly from a chemically doped state. Our results demonstrate the ability to reveal excitation pathways in correlated materials by x-ray spectroscopies, which is relevant for ultrafast materials design. 
\end{abstract} 

\maketitle

Strongly correlated materials host some of the most intriguing states of matter due to the competition between interaction-induced localization and the itinerant nature of electrons, and they are therefore ideal candidates to realize material control on ultrafast timescales \cite{Giannetti2016,delaTorre2022}. Paradigmatic examples are Mott and charge-transfer (CT) insulators, whose optical properties are determined by the charge transfer between the ligand (typically $p$) and the correlated orbital (typically $d$) states \cite{sawatzky1984, zaanen1985, zaanen1986}.  Element- and site-selective information on many-body states in such materials can be obtained by resonant soft x-ray absorption and emission spectroscopies~\cite{de_groot_2p_2021,ament2011,kuo2017challenges} complementing optical spectroscopy~\cite{newman1959,powell1970}. A comparison of core level absorption edges and their fine structure with cluster calculations~\cite{laan1986,laan1988,deGroot1993,Tanaka1994,haverkort2012,sipr2011} or dynamical mean-field theory (DMFT)~\cite{haverkort2014bands,cornaglia2007,hariki2017} can provide detailed information about the CT gap, Coulomb repulsion~\cite{laan1986,okada1997,agrestini2019,burnus2008}, $d$-$d$ multiplet excitations~\cite{rossi2021,ament2011}, and the hybridization between the ligand and correlated orbitals~\cite{schuler2005,luder2017theory}. Furthermore, ligand absorption edges probe the itinerant states~\cite{schuler2005}, whose nature is crucial for understanding the low-energy physics in chemically doped systems~\cite{taguchi2008,kunevs2007,bala1994}. 

Femtosecond time-resolved x-ray spectroscopy is sensitive to low energy excitations and transient energy shifts \cite{stamm_2010,smith_JPCL_2020,Rothenbach2019,diesen_PRL_2021,sidiropoulos_PRX_2021,lojewski_2023}. It is, thus, a potentially powerful tool for investigating photo-excited non-equilibrium states in Mott and CT insulators. Recently, a strong sub-gap excitation has been shown to modify the gap size during the pulse both in cuprate~\cite{baykusheva_2022} and nickelate~\cite{wang2022,granas2022} CT insulators. These effects can be  attributed to photo-manipulations of the screening environment~\cite{baykusheva_2022}, the magnetic order~\cite{wang2022}, or the hybridization between correlated and itinerant orbitals~\cite{granas2022}, coined dynamical Franz-Keldysh effect~\cite{tancogne2020}. A natural open question is how this picture changes in the case of resonant excitations, which create long-lived charge carriers in the conduction band, and lead to so-called photo-doped states which  may host various nontrivial quantum phases \cite{stojchevska2014,li2018,kaneko2019,li2020}. To understand these states, and to potentially design targeted excitation pathways, it is important to clarify the different nature of the charge carriers in photo-doped and chemically doped states, and to disentangle effects such as band shifts due to dynamical screening \cite{golez2019, tancogne2018} and charge redistribution between orbitals \cite{sandri2015,he2016,golez2019,beaulieu2021} from the dynamics of  Mott excitons or electronic $d$-$d$ excitations \cite{Strand2017, Rincon2018, gillmeister2020, aaron2022}. 

\begin{figure}
\includegraphics[width=1.0\linewidth]{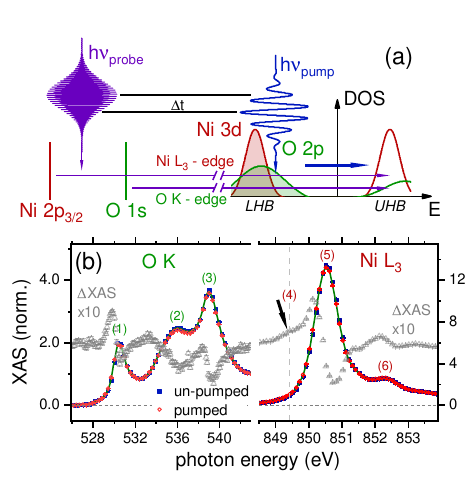}
\caption{(a) Transitions by the optical pump $h\nu_{\mathrm{pump}}$ and the x-ray probe $h\nu_{\mathrm{probe}}$ which is delayed by $\Delta t$. (b) Ground state (blue squares) and pumped (red circles) XAS at the O $K$ (left) and Ni $L_3$ edges (right) for $\Delta t=0.5$~ps and 4~mJ/cm$^2$ fluence. The pumped XAS are modeled based on the static XAS (green line), see \cite{supp}. The gray data show the pump-induced difference $\Delta$XAS vertically offset. }
    \label{fig1}
\end{figure}

In this Letter, we demonstrate for the paradigmatic CT insulator NiO how time-resolved x-ray absorption spectroscopy (tr-XAS) combined with non-equilibrium dynamical mean-field theory (DMFT)~\cite{georges1996,aoki2014} can provide exactly this information upon resonant pumping. We identify energy shifts and lineshape modifications in the excitonic peaks at the Ni $L_3$ and O $K$ absorption edges, and link them to orbital occupations and the screening environment. 
In addition, we resolve a short-lived Ni $L_3$ pre-edge feature that represents many-body multiplets, i.e. $d$-$d$ excitations, as confirmed by the non-thermal generalization of multiplet ligand field theory. 

\begin{figure}
\includegraphics[width=1.0\linewidth]{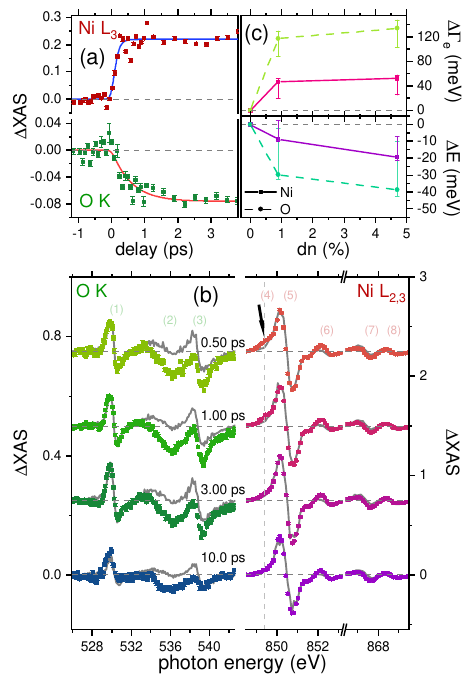}
\caption{(a) Time-dependence of $\Delta$XAS for photon energy 539.15~eV (O $K$) and 850.15~eV (Ni $L_3$), with exponential fits (lines), for the excited charge carrier density $dn=4.7$\% (see \cite{supp} for the determination).
(b) $\Delta$XAS at the O $K$ and Ni $L_3$ edges for 4~mJ/cm$^2$ fluence at the indicated $\Delta t$ and offset vertically. The gray lines show results obtained by a shift $\Delta E$ and broadening $\Delta \Gamma$ of the static spectra, see \cite{supp}. The best fit is used to determine these two parameters.
(c) $\Delta E$ and $\Delta \Gamma$ at the O $K$ and Ni $L_3$ edges from fits in (b) as a function of $dn$ determined at $\Delta t=0.5$~ps. 
}
    \label{fig2}
\end{figure}

Figure~\ref{fig1}(a) shows a sketch of tr-XAS at the Ni $L_3$ ($2p_{3/2} \rightarrow 3d$) and O $K$ ($1s \rightarrow 2p$) edges. The experiments were performed at room temperature at the Spectroscopy and Coherent Scattering (SCS) instrument of European XFEL using a pump-probe setup with an effective time resolution of 80~fs \cite{LeGuyader2023, lojewski_2023, note_data}. Laser pulses with 4.7~eV photon energy, 35~fs pulse duration, and 0.8-4~mJ/cm$^2$ incident fluence were employed to pump 37~nm thick polycrystalline NiO films \cite{supp} above the CT gap. This pumping involves excitations from the O $2p$ states to the upper Hubbard band (UHB) \cite{gillmeister2020}, see Fig.~\ref{fig1}(a). Figure~\ref{fig1}(b) depicts the O $K$ and Ni $L_3$ XAS signals before and after photoexcitation at a pump-probe delay of $\Delta t=0.5$~ps, with the pump-induced difference $\Delta$XAS shown by the gray dots. We find a spectral shift of all features, labeled as (1)-(6), to lower x-ray photon energies, as indicated by the derivative-like shape of $\Delta$XAS. We also measured the time dependence at fixed photon energies at both edges, see Fig.~\ref{fig2}(a). Exponential fits to these transients, convoluted with 80~fs time resolution \cite{lojewski_2023}, indicate a rise time of $99 \pm 20$~fs for 4.0~mJ/cm$^2$ 
pump fluence at the Ni $L_3$ edge and $580 \pm 140$~fs 
at the O $K$ edge. These energy shifts build up, reach a plateau within $2$~ps, and decay on longer timescales. This is in contrast to previous tr-XAS on NiO upon subgap pumping, which showed (smaller) modifications only during the pulse~\cite{wang2022,granas2022}. We attribute the difference to the presence of long-lived holon-doublon pairs after the above-gap excitation. These cannot recombine due to the kinetic constraints~\cite{murakami2023,sensarma2010,lenarcic2013}
and modify the screening environment~\cite{golez2017,golez2015}. 

In a first analysis we compare the spectra of the photo-excited system to curves modelled by a red-shift $\Delta E$ and a broadening $\Delta \Gamma$ of the static spectra, c.f. Fig.~\ref{fig2}(b). In Fig.~\ref{fig2}(c) we plot the best fit $\Delta E$ and $\Delta \Gamma$ at $\Delta t=0.5$~ps as a function of the excitation density $dn$, i.e., the density of pump-excited electrons relative to all valence electrons (see \cite{supp} for the determination of $dn$). The changes in the broadening (or, since $\Gamma \propto 1/\tau$, the changes in the lifetime) are larger for the itinerant $p$ than for the $d$ electrons \cite{lojewski_2023, supp}. Such transient spectral broadening was previously shown to result from lattice excitation \cite{Rothenbach2019, Rothenbach2021}. We thus conclude that lattice excitation, most likely trapping of photo-doped charge carriers by polaron formation~\cite{karsthof2019,ishikawa2017} found on few ps timescales in NiO \cite{biswas2018}, contributes to the long lifetime of the photoexcited state. 

While some parts of the spectrum can be reproduced by the shift and broadening, there are important differences: These include parts of the O $K$ spectrum around $536$ eV, as well as an additional pump-induced feature in the pre-edge region of the Ni $L_3$ edge (see vertical arrow in Fig.~\ref{fig2}(b)), which decays within less than 3~ps. 
 
In the following, we will consider the electronic contribution to the bandgap renormalization and comment on possible polaronic contributiona at the end. The redshift is a generic consequence of photo-doping in CT insulators, which arises simultaneously from (i) dynamical screening of the Coulomb interaction parameters on the transition metal site, and (ii) nonlocal Coulomb interactions between photo-doped ligand holes and electrons in the core and valence orbitals (Hartree shift)~\cite{tancogne2018,tancogne2020,golez2019}. As a first illustration, consider a simple cluster consisting of a valence ($d$), core ($c$), and ligand ($p$) orbital, with density-density Coulomb interactions $U_{dd}$, $U_{cd}$, $U_{pd}$, $U_{cp}$, but no $p$-$d$ hybridization. Upon photo-excitation, the XAS energy for the transition from the core level to the valence $d$ level will shift like $\Delta E_\text{XAS}=  [\Delta U_{dd}+\Delta\epsilon_d-\Delta\epsilon_c -\Delta U_{cd}] + [\Delta N_p(U_{pd}-U_{cp})]$; here the first square bracket represents the change of the local interactions ($U_{dd}$, $U_{cd}$) and level positions ($\epsilon_d$, $\epsilon_c$) by modified screening \cite{tancogne2018,golez2019}, while the second term is a Hartree shift due to the addition of ligand holes ($\Delta  N_p=N_p-N_p^{(0)}$ is the change in the total ligand occupation). 

To demonstrate that such Coulomb shifts prevail beyond the simplistic atomic model, we perform a lattice simulation which includes both a microscopic description of dynamical screening and Hartree shifts. We employ a minimal model for a CT insulator, including one transition metal $d$ orbital and two oxygen $p$ orbitals per unit cell. The hopping between the $d$ and $p$ orbitals $t_{dp}$, the crystal-field splitting $\Delta_{dp}$ and Coulomb interaction parameters are adjusted to match the equilibrium spectral function of NiO and the only free parameter is $U_{cp}$,  which is typically not considered in equilibrium, see \cite{supp} for details. To capture the strong correlations from the Hubbard interaction $U_{dd}$ and the photo-induced changes in screening, we use  the GW+DMFT formalism~\cite{golez2017, golez2019, golez2019a}.  We simulate the photo-excitation by a time-dependent electric field pulse.  The XAS signal is calculated by solving an auxiliary impurity problem that includes  an additional core-level with a lifetime $1/\Gamma$ ($\Gamma=0.05 \text{ fs}^{-1}$); see \cite{werner2022} for details. X-ray energies are measured relative to the position of the main excitonic resonance $|E_c|$ in the equilibrium spectrum. 

\begin{figure}
\includegraphics[width=1.0\linewidth]{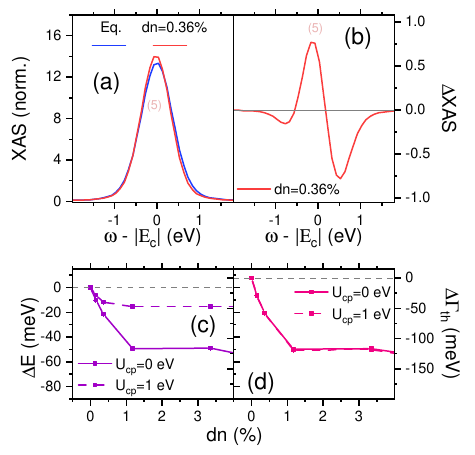}
\caption{(a) Calculated equilibrium XAS (blue) and photo-doped XAS (red line) at photo-doping $dn=0.36$\% and a delay of $\Delta t=7$~fs after the pulse, with the photo-induced change $\Delta$XAS in (b). Shift $\Delta E$~(c) and lifetime change $\Delta \Gamma$~(d) of the exciton peak versus photo-doping $dn$ determined from the calculations for the given values of the core-oxygen interaction $U_{cp}$ and $U_{dd}=U_{cd}=7.5$~eV. 
}
    \label{fig3}
\end{figure}

In the simulation, the photo-excited charge carriers quickly relax to the edge of the charge-transfer gap and are then trapped due to kinetic constraints. 
While the full GW+DMFT simulations are restricted to short times, previous DMFT simulations for a comparable gap demonstrated a long (ps) lifetime of the photo-doped charge carriers \cite{dasari2021}. We therefore expect  that also the XAS signal will not change much on this scale, and can be compared to the long-lived experimental signal in the photo-doped state. The results in Fig.~\ref{fig3} confirm an almost rigid red-shift of the XAS line after photo-doping. The relative importance of the Hartree shift and dynamical screening  does not affect this qualitative behavior, but only the magnitude of the shift (compare in Fig.~\ref{fig3}(c) the results for $U_{cp}=U_{pd}=1$ eV, for which the Hartree shifts in the atomic model cancel, and those for $U_{cp}=0$).  This indicates that the Coulomb shift of the XAS exciton is a generic feature of photo-doped systems, which arises jointly from the dynamical screening of the local interactions and Hartree shifts. Polaronic effects~(stretching of the Ni-O distance)~\cite{ishikawa2017} are further candidates for the bandgap renormalization~\cite{werner2015,erhart2014} and future improvements in the time resolution should enable a separation of the electronic and lattice contributions due to their different timescales.

Our calculations show that the redshift increases linearly with the photo-doping density and then saturates (Fig.~\ref{fig3}c), which is in agreement with the experiment (Fig.~\ref{fig2}(c)). A notable difference to the experiment is that the spectra in the simulations do not broaden but become even more narrow (c.f.~Fig.~\ref{fig1}(a) and Fig.~\ref{fig3}(a)). This may be because the simulation does not capture additional scattering channels, which can lead to a change of the lineshape of the exciton, such as 
lattice~(polaron)~\cite{ishikawa2017} fluctuations.

An interesting question is whether time-resolved XAS can disentangle the different contributions to the Coulomb shift. Assuming a cancellation of the Hartree shifts ($U_{cp}=U_{pd}$) and no screening of $U_{cd}$, $\Delta$XAS would measure the change of the Hubbard interaction $\Delta U_{dd}$, as concluded previously \cite{baykusheva_2022}. However, the observed energy shifts (see Fig.~\ref{fig2}(c)) of less than $100$ meV are small compared to the interaction parameters, so that a statement on the detailed origin of the shift would require precise knowledge of $U_{dd}$, $U_{cd}$, $U_{pd}$ and $U_{cp}$, which is beyond the capability of current first principles approaches such as constrained RPA~\cite{Aryasetiawan2004}. These difficulties prevent a clear quantification of the relative importance of Hartree shifts and dynamical screening and we expect that both contributions are significant. Nevertheless, our analysis proves a high sensitivity of $\Delta$XAS to the microscopic interactions and therefore shows that time-resolved XAS can give information on various Coulomb parameters (and their changes upon photo-doping) that is not easily obtained otherwise, including the nonlocal terms $U_{pd}$ and $U_{cp}$ which are often neglected even in equilibrium \cite{okada1997,de_groot_2p_2021}.

\begin{figure}
\includegraphics[width=1.0\linewidth]{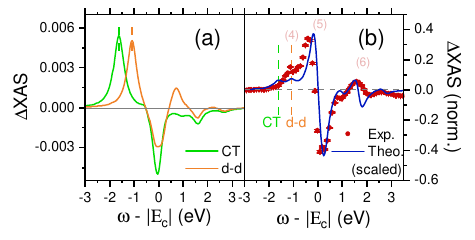}
\caption{(a) Change $\Delta$XAS[$\omega$,$\alpha_a$] in the XAS signal due to the nonthermal population of a low-lying excited state of predominantly $d$-$d$ excitation character ($\alpha_3$=0.02) or charge-transfer~(CT) character ($\alpha_{12}$=0.02). (b) Comparison of $\Delta$XAS between experiment (Ni $L_3$, $\Delta t=0.25$~ps) and theory for
nonthermal orbital populations, 
with the blue curve scaled to match experiment at transient feature (5). The thermal reference spectrum was calculated for $T=100$~K. } 
    \label{fig4}
\end{figure}

Having interpreted the red shift of the $L_3$ edge, we now focus on the transient pre-edge feature in the Ni $L_3$ $\Delta$XAS at $\Delta t=0.25$~ps (Fig.~\ref{fig4}(a)), which is absent at the longest delay times (see Fig.~\ref{fig2}(b)). This motivates an analysis of the multiplet structure of NiO beyond our simple charge-transfer insulator model. Ref.~\cite{werner2022} showed that the positions of the excitonic resonances in the XAS signal of photo-excited systems within time-dependent DMFT can be well reproduced by atomic limit calculations and that their relative amplitudes in the full many-body state on the lattice allow to measure the time-dependent weight of the different local multiplets. This observation motivates us to generalize the multiplet ligand theory~\cite{haverkort2012} to nonthermal distributions~\cite{carneiro2017}. We follow the standard procedure and define an extended Anderson impurity problem, starting from the DFT electronic structure~\cite{haverkort2014bands,luder2017theory}. 
Note that we employ the nonthermal occupations 
$\rho_{ii}[\alpha_a]=\frac{\rho^{th}_{ii}+\alpha_a \delta_{ia}}{\text{Tr}[\rho^{th}_{ii}+\alpha_a \delta_{ia}]}$ with $\alpha_a$ representing the extra weight for the $a$-th eigenstate. These calculations were performed using the EDRIXS library~\cite{wang2019}, include the full $d$ shell as well as the ligand bands, and they allow us to distinguish between photo-induced $d$-$d$ transitions and charge transfer excitations, in contrast to earlier cluster studies~\cite{laan1986,laan1988}. 

We first discuss the photo-induced difference between the nonthermal state and the ground state, $\Delta$XAS[$\omega$,$\alpha_a$]=\text{XAS}($\omega$,$\rho[\alpha_a]$)-XAS($\omega$,$\rho^\text{th}$), see Fig.~\ref{fig4}(a). In the many-body spectrum, there exists a manifold of states around 1~eV above the ground states with a redistribution of electrons within the $d$ shell. We thus attribute them to the $d$-$d$ excitations, see \cite{supp} for a detailed analysis. An excess occupation in these excited states leads to a characteristic modification of the XAS signal, with a pronounced peak appearing below the main resonance~(852 eV), a reduction of the spectral weight at the resonance~(853 eV) and an additional increase around 854 eV. Even higher up in the spectrum, we identify a charge-transfer excitation \cite{supp,agrestini2019,burnus2008}, which results in a photo-induced peak roughly 0.5 eV lower in energy than the $d$-$d$ peak. The latter also leads to a reduction of the spectral weight at the excitonic resonance. For a more direct comparison with the experiment, we can combine the effects of screening and the photo-induced changes from the multiplet ligand field theory. Specifically, we consider a rigid band shift of $\omega_\text{shift}=50$ meV of the XAS signal and add to it the photoinduced changes associated with two states with $d$-$d$ and charge-transfer character,  $\Delta$XAS($\omega$)=XAS($\omega$)-XAS($\omega+\omega_\text{shift}$)+$\sum_{a=3,12}$ $\Delta$XAS[$\omega$,$\rho[\alpha_a]$] where $a=3,12$ refers to the states listed in Tab.~1 of \cite{supp} and the extra weights are set to  $\alpha_a=2\%$. We have checked that different choices for the nonthermal populations lead to qualitatively similar results. The comparison with the experiment shows an excellent agreement in the photo-renormalized XAS signal below and above the main excitonic resonance. 

A comparison of photo- and chemical doping is instructive. Chemical doping leads to the appearance of a pre-edge feature at the O $K$ edge \cite{kuiper1989} while such a pre-edge peak is absent at the Ni $L_{3,2}$ edges \cite{vanelp1991}. Photo-doping produces the opposite result, i.e., the absence of a pre-edge feature at the O $K$ edge (which by comparison to \cite{kuiper1989} would be expected at 527~eV), and the presence of such a feature at the Ni $L_3$ edge. 
The difference between chemically doped and photo-doped spectra is a key observation which shows that photo-doped holes differ in nature from chemically dopes holes: They might occupy different $p$-$d$ hybrid orbitals (such as non-bonding configurations instead of the Zhang-Rice doublet~\cite{bala1994}), or there may exist excitonic correlations between photo-doped electrons and holes.

In conclusion, we showed that photo-doping by above charge-transfer gap excitations in NiO results in characteristic transient signatures in XAS. These are (i) energy shifts of the excitonic peaks of several 10~meV for photo-doping on the order of 1\%, which persist for tens of ps due to the long lifetime of the photo-doped state and are potentially enhanced by polaronic effects~\cite{karsthof2019,biswas2018,carneiro2017}, and (ii) many-body  multiplet excitations in the Ni $L_3$ pre-edge region. 
The good qualitative agreement between the tr-XAS measurements and a simple model for charge-transfer insulators demonstrates that energy shifts resulting from photo-induced changes in the electrostatic Hartree energy are a generic feature in transition metal oxides. Screening-induced changes in the interaction strengths play a role in determining these shifts. 
Our work furthermore establishes that tr-XAS of photo-excited states is sensitive to the interaction between core electrons and ligand holes, which is important for a quantitative interpretation of the spectral changes.  
All these observations reveal generic signatures of charge-transfer insulators, providing a basis for a future systematic analysis of these microscopic phenomena in this important material class, e.g. through the application of sum rules. Disentangling non-equilibrium multiplet effects from time-dependent renormalization of interaction parameters and Hartree shifts further promises to aid the design of tailored excitation protocols for reaching specific photo-doped states.

We acknowledge European XFEL in Schenefeld, Germany, for provision of X-ray free-electron laser beamtime at the SCS instrument and thank the staff for their assistance. Funded by the Deutsche Forschungsgemeinschaft (DFG, German Research Foundation) through  Project No. 278162697 - SFB 1242. UB, PW, and ME acknowledge support from the Deutsche Forschungsgemeinschaft (DFG, German Science Foundation) through FOR 5249-449872909 (Project P6). D.G. acknowledges the support by the program No. P1-0044, No. J1-2455 and MN-0016-106 of the Slovenian Research Agency (ARRS).

T. L. performed the experiments and analyzed the data. D. G. did the calculations. Both contributed equally to this work.


\begin{thebibliography}{75}
\bibitem [{not()}]{note_data}%
  \BibitemOpen
  \href@noop {} {\bibinfo {title} {Data recorded for the experiment at the
  {European XFEL} are available at
  doi:10.22003/xfel.eu-data-002589-00.}}\BibitemShut {Stop}%
\end{thebibliography}

%

\end{document}